\documentstyle[amssymb,twocolumn,prl,aps,epsfig]{revtex}

\begin{document}
\draft
\twocolumn[\hsize\textwidth\columnwidth\hsize\csname @twocolumnfalse\endcsname
\title{Far-infrared transmission studies of $c$-axis oriented superconducting MgB$%
_{2}$ thin film}
\author{J. H. Jung,$^{1,2}$ K. W. Kim,$^{1}$ H. J. Lee,$^{1}$ M. W. Kim,$^{1}$ T. W.
Noh,$^{1}$ W. N. Kang,$^{3}$ Hyeong-Jin Kim,$^{3}$ Eun-Mi Choi,$^{3}$ C. U.
Jung,$^{3}$ and Sung-Ik Lee$^{3}$}
\address{$^{1}$School of Physics and Research Center for Oxide Electronics, Seoul National University, Seoul 151-747, Korea}
\address{$^{2}$Center for Strongly Correlated Material Research, Seoul National University, Seoul 151-747, Korea}
\address{$^{3}$National Creative Research Initiative Center for Superconductivity, Department of Physics, Pohang University of Science and Technology, Pohang 790-784, Korea}
\date{May 4, 2001}
\maketitle

\begin{abstract}
We reported far-infrared transmission measurements on a $c$-axis oriented
superconducting MgB$_{2}$ thin film in the frequency range of 30 $\sim $ 250
cm$^{-1}$. We found that these measurements were sensitive to values of
scattering rate $1/\tau $ and superconducting gap $2\Delta $. By fitting the
experimental transmission spectra at 40 K and below, we obtained $1/\tau =$
(700 $\sim $ 1000) cm$^{-1}$ and $2\Delta (0)\cong $ 42 cm$^{-1}$. These two
quantities suggested that MgB$_{2}$ belong to the dirty limit.
\end{abstract}


%
%\twocolumn[\hsize\textwidth\columnwidth\hsize\csname @twocolumnfalse\endcsname

\vskip1pc] \newpage

The surprising discovery of superconductivity in a known compound MgB$_{2}$
with a high $T_{C}$ of about 39 K has attracted lots of attention from solid
state community and initiated a flurry of activities to understand its
properties\cite{akimitsu}. Most investigations on this binary compound yet
indicate that MgB$_{2}$ behave as a phonon mediated superconductor\cite
{budko,band}. However, there are still little consensus on its important
physical quantities related to electrodynamic and superconducting responses.
Near $T_{C}$, reported values of $dc$ resistivity $\rho _{dc}$ are in the
range from 0.38\cite{canfield} to 75 $\mu \Omega $cm\cite{akimitsu}, and
those of carrier density $n$ vary from 6.7$\times $10$^{22}$\cite{canfield}
to 1.5$\times $10$^{23}$/cm$^{3}$\cite{kang_cm}. Moreover, reported values
of superconducting gap $2\Delta $ also vary from 3\cite{marel} to 16 meV\cite
{chen}. Since such physical quantities are closely related to the nature of
the superconductivity, correct determination of their values cannot be
overemphasized.

It is well known that optical spectroscopy is a powerful method to measure
such important parameters as scattering rate $1/\tau $, plasma frequency, $%
2\Delta $, and coherence effects\cite{IRers}. Since the skin depth of light
is about 1000 \r{A}\ in the far-infrared (IR) region, this technique is able
to provide information on $2\Delta $ complementary to other surface
sensitive techniques, such as tunneling and photoemission measurements.
However, there have been a few reports on the optical properties of MgB$_{2}$%
. Gorshunov {\it et al}.\cite{marel} measured grazing incident reflectivity
of a polycrystalline pellet, and provided a lower estimate of $2\Delta $ to
be 3 $\sim $ 4 meV. On the other hand, Pronin {\it et al}.\cite{pronin}
measured complex optical conductivity using a MgB$_{2}$ thin film in the
frequency range of 4 $\sim $ 30 cm$^{-1}$. For MgB$_{2}$, its far-IR bulk
reflectivity is very close to 1.0 due to the small value of $\rho _{dc}$, so
the commonly used Kramers-Kronig analysis on the reflectivity is difficult
to use and will result in large errors. A transmission measurement using a
superconducting film is a superior method, since it is much more sensitive
to small changes in optical constants. However, the measurements by Pronin 
{\it et al}. were limited below 30 cm$^{-1}$, so characteristic features of
the gap temperature evolution could not be observed.

In this Letter, we investigated electrodynamics of a $c$-axis oriented MgB$%
_{2}$ thin film ($T_{C}$ $\sim $ 33 K) using transmission measurements in
the frequency range of 30 $\sim $ 250 cm$^{-1}$. We found that the
transmission spectra $T(\omega )$ were sensitive to changes in sheet
resistance $R_{\Box }$ and $1/\tau $ of the $ab$-plane. Using the simple
Drude model, we estimated $R_{\Box }=$ 10.3$\pm $0.2 $\Omega /\Box $ and $%
1/\tau =$ 700 $\sim $ 1000 cm$^{-1}$ at 40 K. Below $T_{C}$, a peak due to $%
2\Delta $ appeared and moved to a higher frequency with decreasing
temperature. The $ab$-plane $2\Delta (T)$ seemed to follow the temperature
dependence of the BCS theory and was estimated to be about 42 cm$^{-1}$ (5.2
meV) at 5 K. Comparing the values of $2\Delta (0)$ and $1/\tau $, we
suggested that MgB$_{2}$ should belong to the dirty limit.

A high quality $c$-axis oriented MgB$_{2}$ film was deposited on a Al$_{2}$O$%
_{3}$ substrate using a pulsed laser deposition technique, as already
reported elsewhere\cite{kang_science}. X-ray diffraction measurement showed
that most of grains were oriented with their $c$-axes normal to the
substrate. Using the four-probe method, its $dc$ resistance was measured and
showed a rather sharp $T_{C}$ near 33 K\cite{thick}. Temperature dependent $%
T(\omega )$ were measured with a resolution of 5 cm$^{-1}$ for the frequency
range of 30 $\sim $ 250 cm$^{-1}$ using a Fourier transform
spectrophotometer.

In the thin-film geometry, an approximate form of $T(\omega )$ is well known
for $\lambda \gg \lambda _{p}\gg d$ , where $\lambda $ is the wavelength of
light, $\lambda _{p}$ is the skin depth, and $d$ is the film thickness.
Taking into account of multiple reflections inside the substrate, $T(\omega
) $ can be approximated as

\begin{equation}
T(\omega )=\frac{4n}{|1+\widetilde{N}+\widetilde{y}|^{2}}\times \frac{T_{s%
\text{ }}exp(-\alpha x)}{1-R_{s}R_{f}\text{ }exp(-2\alpha x)},
\end{equation}
where $\widetilde{N}(=n+ik)$, $\alpha (=4\pi \omega k)$, and $x$ are complex
refractive index, absorption coefficient, and thickness of the substrate,
respectively. $T_{s}=4n/|1+\widetilde{N}|^{2}$ and $R_{s}=|1-\widetilde{N}%
|^{2}/|1+\widetilde{N}|^{2}$ are transmission and reflectivity of the
substrate-vacuum boundary, and $R_{f}=|\widetilde{N}-1+\widetilde{y}|^{2}/|%
\widetilde{N}+1+\widetilde{y}|^{2}$ is reflectivity of the substrate-film
boundary with $\widetilde{y}=4\pi \widetilde{\sigma }_{\Box }/c$, where $%
\widetilde{\sigma }_{\Box }=\sigma _{1\Box }+i\sigma _{2\Box }$ is complex
sheet conductance of the film. Using transmission and reflectivity spectra
of Al$_{2}$O$_{3}$, we could obtain $\widetilde{N}$ values independently and
used these to calculate the theoretical $T(\omega )$.

\begin{figure}[tbp]
\epsfig{file=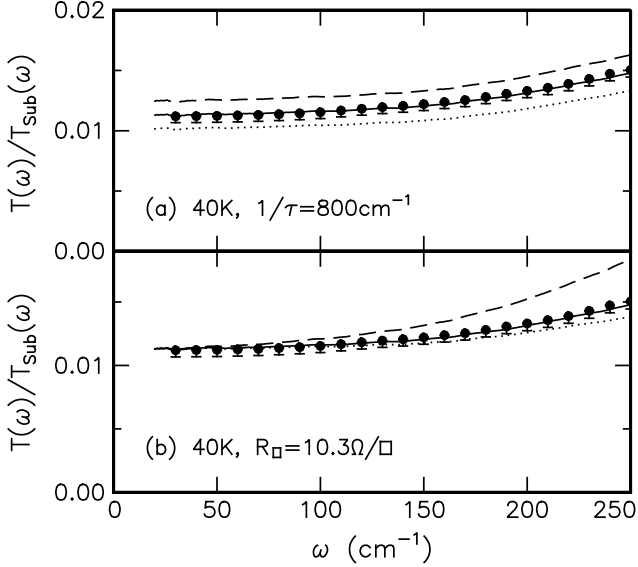,width=3.3in,clip=}
\vspace{2mm}
\caption{$T(\protect\omega )/T_{Sub}(\protect\omega )$ at normal state. In
(a), the dashed, the solid, and the dotted lines represent fitting lines for 
$R_{\Box }=$ 10.8, 10.3, and 9.6 $\Omega /\Box $, respectively. And in (b),
the dashed, the solid, and the dotted lines represent fitting lines for $1/%
\protect\tau =$ 400, 800, and 4000 cm$^{-1}$, respectively.}
\label{Fig:1}
\end{figure}

Far-IR transmission measurements can determine $R_{\Box }$ and $1/\tau $ of
the film quite exactly. Figures 1(a) and 1(b) show the ratio of transmission
spectra between the MgB$_{2}$ film and the Al$_{2}$O$_{3}$ substrate, i.e., $%
T(\omega )/T_{Sub}(\omega )$ at 40 K. To understand effects of $R_{\Box }$,
we calculated $T(\omega )/T_{Sub}(\omega )$ using Eq. (1) with the simple
Drude model. Figure 1(a) shows variations of $T(\omega )/T_{Sub}(\omega )$
with changing $R_{\Box }$ when the value of $1/\tau $ is fixed to be 800 cm$%
^{-1}$: the dashed, the solid, and the dotted lines represent the calculated 
$T(\omega )/T_{Sub}(\omega )$ at $R_{\Box }=$ 10.8, 10.3, and 9.6 $\Omega
/\Box $, respectively. It is clear that $T(\omega )/T_{Sub}(\omega )$ are
quite sensitive to $R_{\Box }$. It was found that $R_{\Box }$ of our film
was 10.3$\pm $0.2 $\Omega /\Box $ at 40 K.

The frequency dependence of $T(\omega )$ should be dependent on $1/\tau $.
At 40 K, $T(\omega )/T_{Sub}(\omega )$ gradually increase as frequency
increases, due to the finite value of $1/\tau $. Figure 1(b) shows
variations of $T(\omega )/T_{Sub}(\omega )$ with changing $1/\tau $ when the
value of $R_{\Box }$ is fixed to be 10.3 $\Omega /\Box $: the dashed, the
solid, and the dotted lines represent the calculated $T(\omega
)/T_{Sub}(\omega )$ at $1/\tau =$ 400, 800, and 4000 cm$^{-1}$,
respectively. It was found that $1/\tau $ of our film was 800$_{-100}^{+200}$
cm$^{-1}$.

\begin{figure}[tbp]
\epsfig{file=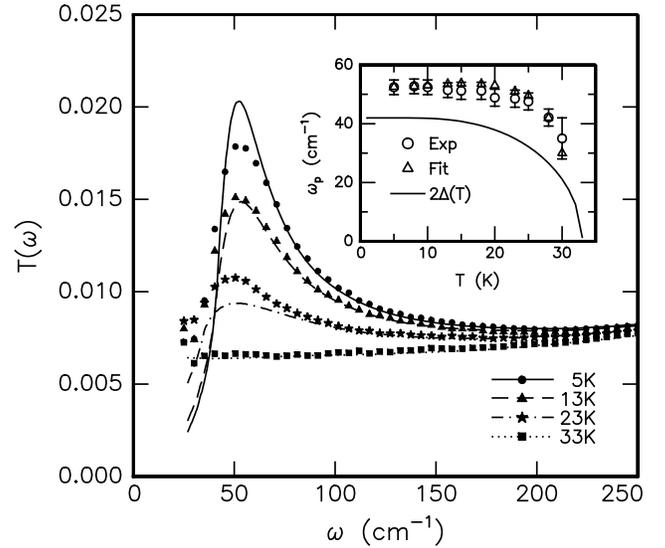,width=3.3in,clip=}
\vspace{2mm}
\caption{Experimental and theoretical results of $T(\protect\omega )$ at
superconducting states. In the inset, the open circles and the open
triangles represent $\protect\omega _{P}$ obtained from the experimental and
the theoretical data, respectively. For comparison, $2\Delta (T)$ is also
shown as a solid line. }
\label{Fig:2}
\end{figure}

Figure 2 shows the temperature dependent $T(\omega )$ of the MgB$_{2}$ film
in its superconducting state, i.e., at temperatures below 33 K. Note that
there is little temperature dependence of $\widetilde{N}$ below 40 K, so $%
T_{Sub}(\omega )$ are nearly flat and independent of temperature\cite{al2o3}%
. In its superconducting state, $T(\omega )$ show a peak-like structure near
50 cm$^{-1}$. As frequency increases, $T(\omega )$ at 5 K gradually increase
up to around 52 cm$^{-1}$ and then approach to those at the normal state. As
temperature increases, the peak height gradually decreases and the peak
position $\omega _{P}$ gradually moves to lower frequencies.

Similar peak structures have been observed for numerous superconducting thin
films, and their $2\Delta $ values have been found to be close to $\omega
_{P}$\cite{cujung}. Such a peak structure can be understood from the complex
optical conductivity spectra $\widetilde{\sigma }(\omega )$. In the normal
state, $\sigma _{1}(\omega )$ at $\omega \ll 1/\tau $ become nearly
frequency independent. In the superconducting state, $\sigma _{1}(\omega )$
below $2\Delta $ become suppressed and the missing spectral weight moves to
the zero frequency to form a delta function, representing a superconducting
condensate. Due to the delta function, $\sigma _{2}(\omega )$\ will have a $%
1/\omega $ dependence. For $\omega \ll 2\Delta $, $T(\omega )$ $\propto $ $%
\omega ^{2}$, which can be easily seen in Eq. (1). On the other hand, for $%
\omega \gg 2\Delta $, $T(\omega )$ in the superconducting state should be
almost the same as those in the normal state. Near $2\Delta $, $\sigma
_{1}(\omega )$ are nearly zero and $\sigma _{2}(\omega )$ drastically
decrease, resulting in a minimum value of the denominator in Eq. (1). So,
there is a peak in transmission.

One of important physical quantities in a superconductor is the ratio
between $1/\tau $ and $2\Delta (0)$, which will determine electrodynamic
responses\cite{tinkham}. For most metal superconductors, such as Al and Pb, $%
(1/\tau )/2\Delta (0)\gtrsim $ 100. For such BCS superconductors in the
extremely dirty limit, their optical responses can be explained by the
Mattis-Bardeen theory\cite{mb}. However, in high temperature
superconductors, their $2\Delta (0)$ values are much larger than those of
metal superconductors, and they are believed to belong to the clean limit,
where $(1/\tau )/2\Delta (0)\sim $ 1\cite{review}. Optical properties of
superconductors in the clean limit are different from those in the dirty
limit: for example, $\sigma _{1}^{s}(\omega )/\sigma _{1}^{n}(\omega )$ in
the clean limit has a steeper rise near $2\Delta $\cite{tinkham}, where $%
\sigma _{1}^{s}(\omega )$ and $\sigma _{1}^{n}(\omega )$ represent $\sigma
_{1}(\omega )$ at superconducting and normal states, respectively. Zimmerman 
{\it et al.} calculated optical conductivity of a homogeneous BCS
superconductor with arbitrary purity\cite{zimmerman}.

In order to explain the temperature dependent $T(\omega )$, we applied the
formula developed by Zimmerman {\it et al.}\cite{zimmerman} with the
measured values of $1/\tau $ and $R_{\Box }$ at 40 K, i.e., 800 cm$^{-1}$
and 10.3 $\Omega /\Box $, respectively. As shown in Fig. 2, $\omega _{P}$ of
the 5 K data could be fitted quite well with $2\Delta (0)\cong $ 42 cm$^{-1}$%
. Note that the $2\Delta (0)$ value is smaller than the $\omega _{P}$ value
by about 10 cm$^{-1}$. Assuming that the temperature dependence of $2\Delta
(T)$ follow the BCS prediction, we calculated $T(\omega )$. The solid
circles (solid line), the solid triangles (dashed line), the solid stars
(dot-dashed line), and solid squares (dotted line) represent the
experimental (calculated) $T(\omega )$ at 5, 13, 23, and 33 K, respectively.
The predicted temperature dependences of $\omega _{P}$ and $2\Delta (T)$ are
shown in the inset as the open triangles and the solid line, respectively.
Within the experimental error bars, $2\Delta (T)$ seems to follow the
prediction of the BCS theory quite well.

Although many experimental studies have been performed on the $2\Delta (0)$
value of MgB$_{2}$, there is little consensus on its magnitude and symmetry.
In specific heat measurements, a couple of groups reported that $2\Delta
(0)/k_{B}T_{C}$ $\sim $ 2.4\cite{walti} or 4.2\cite{kremer}, both of which
were explained in terms of the conventional $s$-wave type BCS model. Using
the same technique, Wang {\it et al.}\cite{wang} reported that the values of 
$2\Delta (0)/k_{B}T_{C}$ were varied form 1.2 to 4.2 and suggested a $d$%
-wave superconductor with nodes in the gap. In photoemission measurements,
Takahashi {\it et al}.\cite{takahashi} found that the superconducting gap
was $s$-like with $2\Delta (0)/k_{B}T_{C}$ $\sim $ 3.0, but Tsuda {\it et al.%
}\cite{tsuda} reported a spectroscopic evidence for two gaps with $2\Delta
(0)$ $=$ 3.4 and 11.2 meV. More seriously, in tunneling experiments, the
values of $2\Delta (0)$ were varied from 4 to 16 meV, and both isotropic and
anisotropic gap symmetries were suggested\cite{stm}. To explain the large
variations of $2\Delta (0)$ values, Hass and Maki\cite{maki} recently
proposed a model of anisotropic $s$-wave superconductivity.

Our measured value of $2\Delta (0)$, i.e., about 42 cm$^{-1}$ (5.2 meV) is
quite smaller than the BCS prediction for the isotropic $s$-wave
superconductor. Since our $c$-axis oriented film has $T_{C}\sim $ 33 K, $%
2\Delta (0)/k_{B}T_{C}$ can be evaluated to be about 1.8. There are at least
three possibilities to explain the small value of $2\Delta (0)/k_{B}T_{C}$.
The first possibility is an existence of dead layers at the film surface and
the film/substrate interface. We found that the reflectivity spectra in the
visible region for numerous thick films varied from film to film, indicating
possible existence of dead layers. Since our film used in this transmission
study is only 500$\pm $70 \r{A} thick\cite{thickness}, effects of the dead
layers could be important. However, the temperature dependent changes in $%
T(\omega )$ are quite large and agree well with the predictions of the BCS
theory, so the dead layer effect cannot be very large. And the dead layer
effect will be less effective for transmission measurements than
reflectivity measurements. The second possibility is inhomogeneities of our
films, especially in $T_{C}$ and/or $2\Delta (0)$. However, our $T(\omega )$
could not be explained by introducing a distribution of $2\Delta (0)$ at a
higher frequency region in our calculation. The last possibility is the
anisotropic gap symmetry. Since our measurements were made on the $c$-axis
oriented film, our $2\Delta (0)/k_{B}T_{C}$ value represents the $ab$-plane
property. The anisotropic $s$-wave superconductivity, suggested by Hass and
Maki\cite{maki}, is consistent with the small $ab$-plane value of $2\Delta
(0)$. \ In order to prove this gap symmetry, it would be useful to perform
polarization dependent optical measurements on a single crystal or an
epitaxial film with its $c$-axis parallel to a substrate.

\begin{figure}[tbp]
\epsfig{file=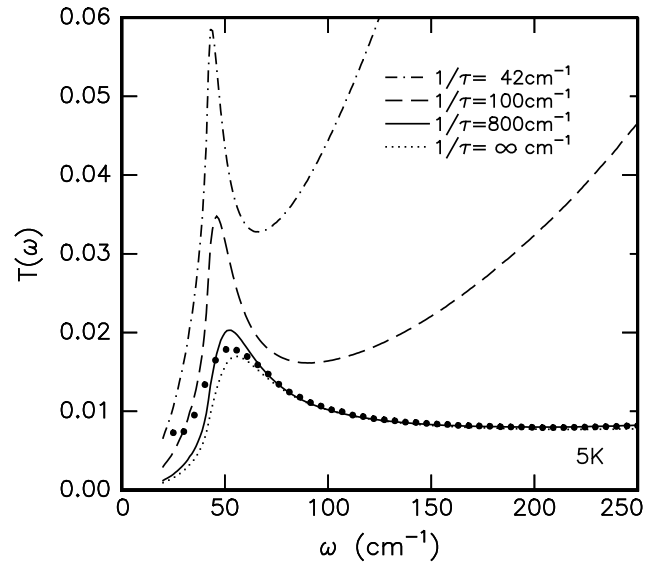,width=3.3in,clip=}
\vspace{2mm}
\caption{Calculated $T(\protect\omega )$ at 5 K with changing $1/\protect\tau
$. The solid circles represent the experimental data. The dot-dashed, the
dashed, the solid, and the dotted lines represent $T(\protect\omega )$ at $1/%
\protect\tau =$ 42, 100, 800, and $\infty $ cm$^{-1}$, respectively. }
\label{Fig:3}
\end{figure}

Many workers claimed that the MgB$_{2}$ superconductor should be in the
clean limit (i.e., $2\Delta (0)\geq 1/\tau $). Canfield {\it et al.}\cite
{canfield} estimated Fermi velocity $v_{F}=$ 4.8$\times $10$^{7}$ cm/s and
mean free path $l=$ 600 \r{A}. Comparing with the superconducting coherence
length $\xi _{0}\sim $ 52 \r{A}\cite{finnemore}, they suggested that MgB$%
_{2} $ should belong to the clean limit. By the relation of $l=v_{F}\times
\tau $, we estimated the value of $1/\tau $ $\simeq $ 42 cm$^{-1}$. However,
our experimental data in the normal state as well as the superconducting
states showed that $1/\tau \approx $ 800 cm$^{-1}$\cite{comment}. We
calculated $T(\omega )$ at 5 K with various values of $1/\tau $, shown in
Fig. 3. Note that the calculated $T(\omega )$ with $1/\tau $ $\simeq $ 42 cm$%
^{-1}$ predict a very sharp peak structure near $2\Delta (0)$ and a steep
increase of $T(\omega )$ at the high frequency region, which do not agree
with our experimental observations. Our experimental $T(\omega )$ seem to be
quite close to predictions of the Mattis-Bardeen theory in the extremely
dirty limit, i.e., $1/\tau =\infty $ cm$^{-1}$. Although the Mattis-Bardeen
theory is slightly better to explain the 5 K data than the Zimmerman's
formula with $1/\tau =$ 800 cm$^{-1}$, the temperature dependence of the
peak can be explained better by the latter method. \ Moreover, our 40 K
transmission data suggested that $1/\tau \approx $ 800 cm$^{-1}$.

In literature, there is a large variation of $\rho _{dc}$ value. Our
measured value of $1/\tau \approx $ 800 cm$^{-1}$ provides a certain
limitation to electrodynamic quantities. Using the reported values of $%
v_{F}= $ 4.8$\times $10$^{7}$ cm/s and $n=$ 6.7$\times $10$^{22}$ /cm$^{3}$%
\cite{canfield}, we estimated values of $l$ and $\rho _{dc}$ were 32 \r{A}
and 24 $\mu \Omega $cm, respectively. Note that, the estimated value of $l$
is somewhat smaller than $\xi _{0}\sim $ 52 \r{A} and that of $\rho _{dc}$
is much larger than 0.38 $\mu \Omega $cm reported for MgB$_{2}$ wire\cite
{t_level}. It is clear that more investigations are necessary to determine
even simple physical quantities, such as $\rho _{dc}$.

In summary, we investigated the $c$-axis oriented superconducting MgB$_{2}$
thin film using transmission measurements. By fitting transmission spectra
at normal and superconducting states, we found that the scattering rate was
(700 $\sim $ 1000) cm$^{-1}$ at 40 K and zero temperature superconducting
gap was about 42 cm$^{-1}$. These electrodynamics quantities suggested that
MgB$_{2}$ should belong to the dirty limit.

This work at SNU and POSTECH is supported by the Ministry of Science and
Technology of Korea through the Creative Research Initiative Program.


\begin{references}
\bibitem{thick}  When we deposited 3000 \r{A} thick films using our film
fabrication technique, we could prepare high quality $c$-axis oriented MgB$%
_{2}$ films with $T_{C}\sim $ 39 K [10]. However, their transmission were
too small to be measured.
\bibitem{thickness}  When the film was assumed to be homogeneous, the film
thickness $d$ were estimated to be 500$\pm $70 \r{A} by spectroscopic
ellipsometry measurement. However, due to the possible existence of dead
layers, we did not use this value to evaluate the $dc$ resistivity of our
film.
\bibitem{t_level}  If we used the values of $\rho _{dc}$ $=$ 0.38 $\mu
\Omega $cm, we found that $d$ should be less than 4 \r{A} to fit the
experimental $T(\omega )$.
\end{references}
\end{document}